\documentstyle[aabib99,amssymb,psfig,epsfig]{l-aa}
 
\begin{document}

\thesaurus{06(02.08.01; 08.02.1; 08.14.1; 13.07.1)}

\title{Merging neutron stars: asymmetric systems}

\author{S. Rosswog \inst{1}
\and M.B. Davies   \inst{2}
\and F.-K. Thielemann  \inst{3}
\and T. Piran  \inst{4}}

\institute{Center for parallel computing (ZPR/ZAIK), Universit\"at zu K\"oln, 
            Germany
	   \and Dept. Physics and Astronomy, University of Leicester LEI 7RH, 
            UK
           \and Departement f\"ur Physik und Astronomie, Universit\"at Basel, 
            Switzerland
           \and Racah Institute for Physics, Hebrew University, Jerusalem, 
		Israel}

\offprints{S. Rosswog, \\ email: rosswog@zpr.uni-koeln.de, \\
                          Fax:   ++49/221/470-5160}

\date{Received ; accepted }

\maketitle

\begin{abstract}

We present the results of 3D, Newtonian hydrodynamic calculations
of the last stages of the inspiral and the final coalescence of neutron star
binary systems. Our focus is on slightly asymmetric systems, where the 
asymmetry stems from either different masses (1.3 and 1.4 ${\rm M}_{\odot}$) or
spins of both components.
Almost immediately after contact a fast rotating, very massive central object
forms. All calculations exhibit baryonic masses above 2.3 ${\rm M}_{\odot}$,
thus based on our calculations it is not possible to decide on the fate of 
the central core of the merged configuration. It might collapse immediately
to a black hole, but also the creation of a supermassive neutron star with
$\sim 2.8$ ${\rm M}_{\odot}$ cannot firmly be excluded. Depending on the 
asymmetry of the system the central object receives a kick of several hundred 
kilometers per second.
Different spins of both components do not jeopardize the formation of (to 
within numerical resolution) baryon free funnels above the poles of the 
central objects. In the case of different masses the less massive components
 get disrupted and engulf the more massive companions that stay 
rather unaffected by the collision.
The amount of ejected material is in a similar range as for symmetric systems 
and could contribute substantially to the enrichment of the Galaxy with 
heavy r-process elements. Test calculations indicate that the amount of 
ejected material is basically determined by the high density behaviour of the 
nuclear equation of state.\\
Test calculations for the hybrid artificial viscosity scheme that is used for
this work are given in the appendix.

\end{abstract}


\section{Introduction}

The coalescence of compact binary objects  has received 
such attention since its inspiral is expected to emit gravitational waves
in the frequency band that is best accessible to ground based gravitational
wave detectors such as LIGO \cite{abramovici92}, VIRGO \cite{bradaschia90},
 TAMA \cite{kuroda97} and GEO600 \cite{luck97}.\\
In addition this scenario has for many years been ``the'' model for the 
central engine
to power gamma ray bursts (GRBs), especially since the final detection of 
counterparts in wavelength
 regimes different from gamma rays 
\cite{vanparadijs97,sahu97,galama97,djorgovski97,costa97,frail97}.
 However, with the BeppoSAX detection 
of GRBs with enormous redshifts  the situation somehow has
 changed: even the coalescence of a neutron star binary with its enormous 
gravitational binding energy of $\sim 5\cdot 10^{53}$ erg might not be able 
to produce $\sim  3 \cdot 10^{54} \cdot (\frac{\Omega}{4 \pi})$ erg in 
gamma rays ($\Omega$ is the beaming angle) that are required for the recent
 burst GRB990123 \cite{blandford99}.
Now there is growing consensus that the bimodal distribution of burst 
durations has to be attributed to two different kinds of progenitors. 
The short ones, that are observationally rather unconstrained due to the 
BeppoSAX trigger time of five seconds,
 possibly emerge from accretion disks around black holes resulting from 
the coalescences of neutron star - neutron star (ns-ns) or low mass black hole
 - neutron star (bh-ns) systems and the longer ones
from the so-called ``failed supernovae'' \cite{bodenheimer83,woosley93}, 
``collapsars'' or ``hypernovae'' \cite{paczynski98,galama98,iwamoto98,wheeler98}. Recent simulations \cite{macfadyen99} indicate that in this case indeed
 a highly energetic jet 
might be driven through the mantle of the collapsing high-mass star.\\
Further interest in the merging scenario arises from its possible importance
for r-process nucleosynthesis 
(Lattimer and Schramm 1974, 1976; Symbalisty and Schramm 1982; Eichler et al.
1989; Meyer 1989; Rosswog et al. 1999, hereafter paper I). 
Despite a reasonable understanding of the underlying nuclear
 processes and many years of intense research, it has not yet been possible to
identify the corresponding production site unambiguously. The ``classical''
 r-process site, type II supernovae, seem not to be able to provide
the entropies required for a reproduction of the solar r-process pattern 
\cite{freiburghaus97,freiburghaus99a,meyer97c} unless very special 
neutrino properties are chosen \cite{mclaughlin99}. The decompression of 
low entropy, low $Y_{e}$ material could be an alternative or supplementary 
scenario.\\
The evolution of the last inspiral stages and the subsequent coalescence 
of close compact binaries have been discussed by a number 
of groups. The first 3D hydrodynamic calculations have been performed by 
Nakamura and collaborators (see Shibata \cite*{shibata93} and
 references therein). Similar 
calculations have been performed by Davies et al. \cite*{davies94}
 who discussed a 
number of physical processes connected with the merging event. Zhuge et al.
 \cite*{zhuge94,zhuge96} focussed in their work on the emission of 
gravitational radiation. Lai, Rasio and Shapiro analyzed close 
binary systems both analytically 
\cite{lai93a,lai93b,lai94a,lai94b,lai94c,lai94d} as 
well as with numerical calculations \cite{rasio92,rasio94,rasio95}
where their main interest was focussed on stability issues.
Many of their results were confirmed  by the work of New and Tholine 
\cite*{new97}.
Ruffert et al.\cite*{ruffert96,ruffert97a} were the first to include 
microphysics (realistic equation of state, neutrino emission) in their
 hydrodynamic calculations.
Rosswog et al. \cite*{rosswog99} focussed in their work on the mass loss during the 
merger event and possible implications for the nucleosynthesis of heavy 
elements.
Recently the interaction of low mass black holes with neutron stars has 
been investigated \cite{ruffert99,kluzniak98b,janka98}.\\
Several attempts have been made to include effects of general relativity (GR)
in approximative ways, in both analytical 
\cite{lai96,taniguchi96,lai97,lombardi97,taniguchi97,shibata97} 
and numerical treatments 
\cite{wilson95,shibata96,wilson96,mathews97,baumgarte97,baumgarte98a,baumgarte98b,shibata98,bonazzola99}. However, the corresponding results
 are still not free of controversies. Recently attempts have been made to give
an SPH-formulation of the Post-Newtonian formalism of Blanchet, Damour and 
Sch\"afer \cite*{blanchet90} \cite{ayal99,faber99}.\\
After having concentrated in previous simulations on symmetric neutron star 
binary systems we want to focus in this investigation on slightly 
asymmetric neutron star binaries. Neutron star systems with mass ratio 
$q\ne 1$ have previously been analyzed by Rasio and Shapiro \cite*{rasio94}
 and Zhuge et al. \cite*{zhuge96}. However, both groups used simple 
polytropes to approximate the equation of state, whereas in the work reported
here we use the realistic nuclear equation of state of Lattimer and 
Swesty (1991).\\
The ingredients of our model are described in Sect. 2, results 
concerning morphology, mass distribution, kick velocities of the central 
objects, possible implications for GRBs, temperatures and ejecta may be found
 in Sect. 3. A summary and discussion of our results are given in Sect. 4.
	
\section{The model}

\begin{table*}[ht]
\caption{Summary of the different runs: 
{\em spin}: 1: no spin + spin with orbit;
2:  no spin + spin against orbit;
3:  no spin + spin in orbital plane 1;
4:  corotation;
5:  no stellar spins;
6:  spins against orbit (Fig. \ref{spins}).
Masses are given in solar units. LS: Lattimer \& Swesty (1991). $\Gamma_{1}
$ and $\Gamma_{2}$ refer to the adiabatic index of the pseudo-polytropic EOS 
given in Eq. (\ref{adindex}). AV refers to the artificial viscosity scheme,
 $E_{s.o.}$  to the total energy at the time
when the backreaction force is switched off, $E_{fin}$ at the end of the 
calculation.}
\begin{flushleft}
\begin{tabular}{ccccccccccc} \hline \noalign{\smallskip}
run & spin & ${\rm M_{1}}$ & ${\rm M_{2}}$ & \# part. & $a_{0}$ [km] 
& EOS & initial equilibrium & AV & $|E_{s.o.}-E_{fin}|/|E_{fin}|$\\ \hline \\
A   & 1 & 1.4 & 1.4 &  20000 & 45 & LS & no & hybrid & 3.1 $\cdot 10^{-4}$\\ 
B   & 2 & 1.4 & 1.4 &  20000 & 45 & LS & no & hybrid & 6.0 $\cdot 10^{-4}$ \\
C   & 3 & 1.4 & 1.4 &  20000 & 45 & LS & no & hybrid & 1.3 $\cdot 10^{-3}$\\
D   & 4 & 1.3 & 1.4 &  20000 & 45 & LS & yes& hybrid &9.8 $\cdot 10^{-4}$ \\
E   & 5 & 1.3 & 1.4 &  20000 & 45 & LS & no & hybrid &1.1 $\cdot 10^{-4}$\\
F   & 6 & 1.3 & 1.4 &  20852 & 45 & LS & no & hybrid &9.0 $\cdot 10^{-4}$\\
G   & 4 & 1.6 & 1.6 &  20974 & 45 & $\Gamma_{1}=2.0, \Gamma_{2}=2.6$& no &
 standard & 5.9 $\cdot 10^{-4}$\\
H   & 4 & 1.6 & 1.6 &  20000 & 45 & $\Gamma_{1}=2.6, \Gamma_{2}=2.0$& no &
 standard &2.6 $\cdot 10^{-3}$\\
I   & 5 & 1.3 & 1.4 & 20000 & 45  & LS & no & standard & 3.1 $\cdot 10^{-4}$ &\\
J   & 5 & 1.3 & 1.4 & 20000 & 45  & LS & no & no & 1.3 $\cdot 10^{-3}$ &\\
\end{tabular}
\end{flushleft}
\label{runs}
\end{table*}

\begin{table*}[ht]
\caption{Masses of the different morphological regions ('co' stands for 
central object, 'ld' for the low density parts outside the disk), 
$ \tilde{{\rm M}}_{\rm co}$ is a lower limit on the gravitational 
mass (see Eq. (\ref{gravmass})) and $a$ is the relativistic stability 
parameter (see text).}
\begin{flushleft}
\begin{tabular}{cccccccc} \hline \noalign{\smallskip}
run & ${\rm M}_{\rm co} \; [{\rm M}_{\odot}]$ & $ \tilde{{\rm M}}_{\rm co} \; [{\rm M}_{\odot}]$  & a & ${\rm M}_{\rm disk} \; [{\rm M}_{\odot}]$ & ${\rm M}_{\rm ld} \; [{\rm M}_{\odot}]$ \\ \hline \\
A & 2.55 & 2.13 & 0.61 & 0.20 & 0.05 \\ 
B & 2.69 & 2.22 & 0.64 & 0.10 & 0.01 \\ 
C & 2.60 & 2.17 & 0.60 & 0.14 & 0.06 \\ 
D & 2.37 & 2.00 & 0.55 & 0.23 & 0.10 \\ 
E & 2.49 & 2.08 & 0.59 & 0.16 & 0.05 \\
F & 2.65 & 2.19 & 0.66 & 0.05 & 3 $\cdot 10^{-3}$ \\ 
\end{tabular}
\end{flushleft}
\label{masses}
\end{table*}

\subsection{The numerical method}

We solve the dynamic equations of fluid motion  in three dimensions 
 using the smoothed particle hydrodynamics (SPH) method. 
Due to its Lagrangian nature this method is perfectly suited to tackling this 
intrinsically three dimensional problem. It is not restricted to a 
computational domain imposed by a grid and it is easy to track the 
history of chosen blobs of matter (e.g. ejected material).
Since this method has been discussed extensively, we restrict ourselves 
here to mentioning just the basic ingredients of our code and refer to the 
literature for further details (see e.g. Hernquist and Katz 
\cite*{hernquist89}, Benz \cite*{benz90a}
and Monaghan \cite*{monaghan92}).\\
The Newtonian gravitational forces are evaluated using a 
hierarchical binary tree as described in Benz \cite*{benz90b}. The 
additional forces arising from the emission of gravitational waves will 
be discussed below. 

\subsection{Gravitational radiation backreaction}

Since the forces emerging from the emission of gravitational waves 
tend to circularize binary orbits, we start our calculations with 
 circular orbits \cite{peters63,peters64} and add the 
radial velocity of a point-mass 
binary in quadrupole approximation (see e.g. Shapiro \& Teukolsky 
\cite*{shapiro83}):
\begin{equation}
\frac{da}{dt}= -\eta a, \quad \eta= \frac{64}{5} \frac{G^{3}}{c^{5}} 
\frac{M^{2} \mu}{a^{4}},
\end{equation} 
where $a$ is the distance between 
the binary components, $M$ the total and $\mu$ the reduced mass of the system.
 In addition to the accelerations from hydrodynamic and Newtonian 
gravitational forces we apply to each SPH-particle in star $i$ the 
backreaction acceleration $\vec{a}_{gwb,i}$ found for a point-mass binary with unequal masses $M_1$ and $M_2$. Starting from
the point mass formulae for the loss of energy $E$ and angular momentum 
$\vec{J}$
due to the emission of gravitational waves on circular orbits, 
$\frac{dE}{dt}= \eta E$ and $\frac{d \vec{J}}{dt}= -\frac{\eta}{2} \vec{J}$,
 and assuming $\frac{dE}{dt}= \sum_i m_i \vec{v}_i \vec{a}_{gwb,i}$ and 
$\frac{d \vec{J}}{dt}= \sum_i m_i \vec{x}_i \times \vec{a}_{gwb,i}$, we derive
\begin{eqnarray}
\ddot{x}_{i,{\rm gwb}}= (-1)^{i+1} \frac{\eta}{M_{i} \; (\vec{r_{12}} \cdot 
\vec{v_{12}})} \left(E x_{12} + \frac{ J \; \dot{y}_{12}}{2}\right)
\label{gwback1}\\
\vspace{0.8 cm}
\ddot{y}_{i,{\rm gwb}}= (-1)^{i+1} \frac{\eta}{M_{i} \;(\vec{r_{12}} \cdot 
 \vec{v_{12}})} \left(E y_{12} - \frac{ J \; \dot{x}_{12}}{2}\right)
\label{gwback2}.
\end{eqnarray}
Here  $\vec{r_{12}}= \vec{r_{1}}-\vec{r_{2}}$, 
$\vec{v_{12}}= \vec{v_{1}}-\vec{v_{2}}$ and their 
components are $x_{12},y_{12};\dot{x}_{12},\dot{y}_{12}$.
For systems with equal masses this reduces to the formula given in
\cite{davies94}. The backreaction acceleration is switched off when
the stars come into contact and the point mass limit is definitely 
inappropriate. In test calculations our backreaction force has been compared
to the frictional force of \cite{zhuge96}, where the accelerations were
calculated according to 
\begin{equation}
\vec{a}_{gwb,1}= -\frac{G}{2} \frac{M_1}{M_2} \frac{\mu}{a} \eta 
\frac{\vec{v}_1}{|\vec{v}_1|}
\end{equation}
for star 1 and similar for star 2. The results found for both 
implementations where almost indistinguishable. A further discussion 
concerning the appropriateness of this approach may be found in 
paper I. 

\subsection{Artificial Viscosity}

The probably-superfluid, almost inviscid neutron star material poses 
a severe problem for a numerical simulation. In addition to the numerical
dissipation arising from discretisation, SPH uses an artificial viscosity 
(AV) to resolve 
shocks properly. The standard form of AV \cite{monaghan92} is known 
to yield good results in pure shocks but to introduce spurious entropy 
generation in shear flows
(for an extensive report on the effects of AV see \cite{lombardi99}).
 In this work a new, hybrid scheme of AV is
used that profits from two recently suggested improvements: the decrease
of the viscosity parameters to low minimum values in the absence of
shocks \cite{morris97} and the so-called Balsara-switch 
\cite{balsara95} to reduce viscosity in pure shear flows. Details and test 
calculations of the new hybrid scheme are given in the appendix.

\subsection{The nuclear equation of state}

The equation of state (EOS) of the neutron star material is an 
important ingredient of the model, but unfortunately it is only poorly known 
and thus introduces large uncertainties into the calculation. In paper I
we found, for example, that the amount of ejected mass
 is very sensitive to the stiffness, i.e. the adiabatic exponent, of the EOS.\\
To describe the macroscopic properties of the neutron stars we use  for 
our calculations the temperature dependent  equation of state of Lattimer
 and Swesty \cite*{lattimer91} in a tabular form. 
This EOS models the hadronic matter 
 as a mixture of a heavy nucleus (representative for an ensemble of heavy 
nuclei), alpha particles (representing light nuclei) and nucleons outside 
nuclei. To mimic the softening of the EOS at high densities resulting from
 the appearance of ``exotic'' particles (such as hyperons, kaons and pions),
 which are neglected in the approach of Lattimer and Swesty, we use the 
lowest available compressibility, $K= 180$ MeV. Technical details concerning 
the EOS may be found in Appendix B of paper I.

\subsection{Masses}

 All known  neutron star binary  systems show a remarkably
small mass difference between the two components
 (e.g. Thorsett \cite*{thorsett96}; Thorsett and Chakrabarti 
\cite*{thorsett99}).
 The maximum known mass difference is $\approx 4
 \%$ for PSR 1913+16. However, the overall dynamics of the merging event is 
rather sensitive to those mass differences (see e.g. Rasio and Shapiro 
\cite*{rasio94}) and
larger asymmetries in the masses cannot be excluded on grounds of the five 
known systems. \\
In the cases where we are interested in the question of how unequal neutron 
star spins alter previous results, we study binary systems where both 
components have a baryonic mass of 1.4 $ {\rm M}_{\odot}$ (for details see 
Table \ref{runs}).
To examine the effects of unequal masses we consider systems containing a 
$1.3 {\rm M}_{\odot}$ and a $1.4 {\rm M}_{\odot}$ star. Keeping in mind that
the well-known systems are centered around 1.35 $ {\rm M}_{\odot}$ this choice
is highly realistic.

\subsection{Stellar spins}

\begin{figure}[t]
\psfig{file=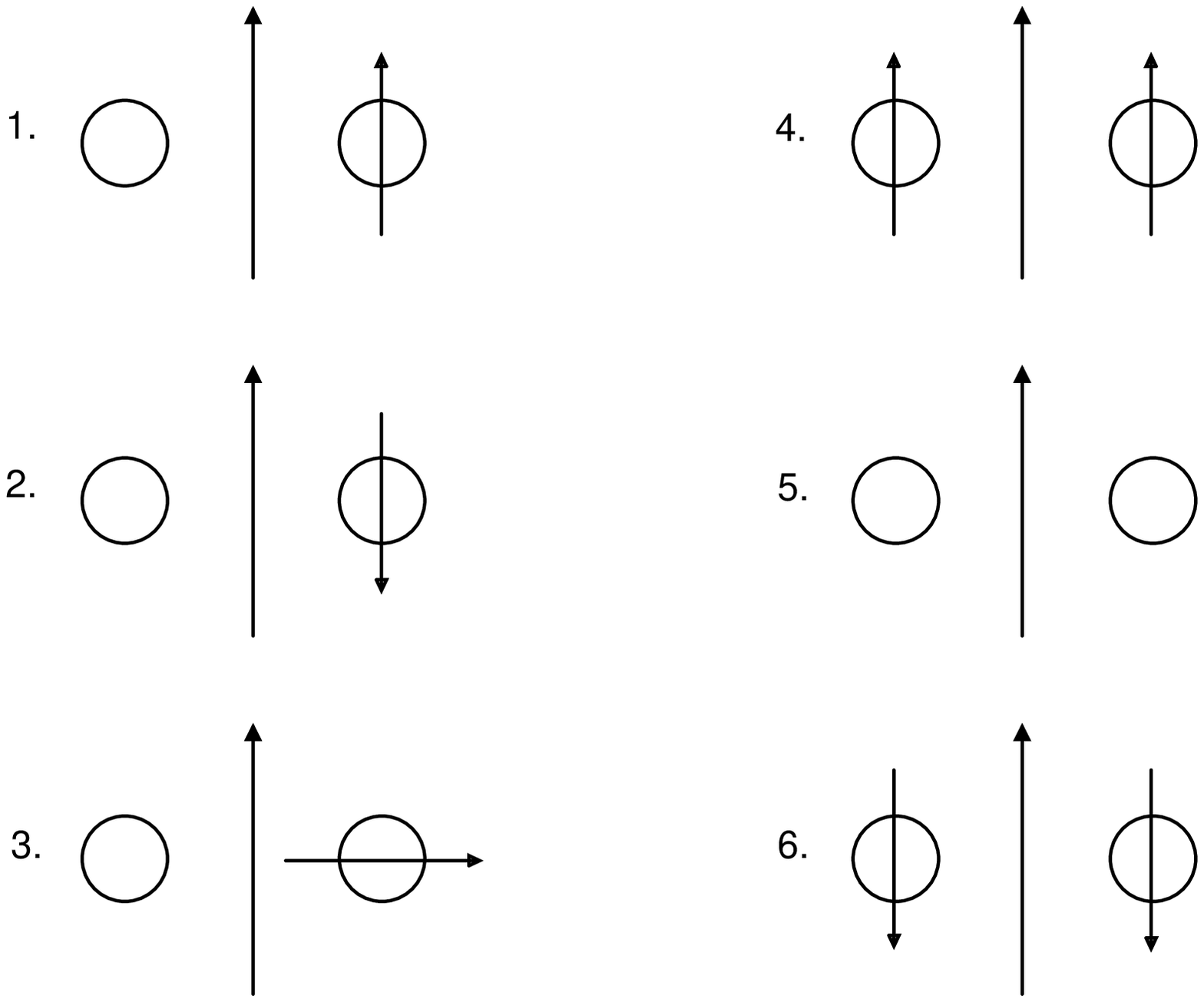,width=12cm,angle=-90}
\caption{\label{spins}The investigated spin orientations. The large arrow 
symbolizes the orbital and the small one the spin angular momentum.}
\end{figure}

Bildsten and Cutler \cite*{bildsten92} and Kochanek \cite*{kochanek92} have 
shown that the internal 
viscosity of the neutron star material is by far too low for a tidal 
locking of both components within the short time scale during which the stars
 can tidally interact. They concluded that the system should be close to 
irrotational, i.e. spin angular momentum should be negligible in comparison
 to the orbital angular momentum. Observationally there is not very much
 known about spin distributions in neutron star binary systems. 
We regard it an interesting question whether the amount of ejected mass 
is altered decisively by unequal stellar spins.
The tremendous interest in the coalescence scenario of two neutron stars
has been triggered -among other reasons- by its possible connection to
 the still poorly understood GRBs. After a decade as {\em the} model for
the central engine of GRBs it is currently favoured as the most promising
model for (at least) the subclass of short bursts.
One of the features that makes this scenario appealing is the emergence 
of baryon 
free funnels above the polar caps of the central object, which have been found
in the calculations of symmetric systems \cite{davies94,ruffert97a,rosswog99}.
 These funnels are attractive sites for
 the fireball scenario since they are close to the central source but 
relatively free of baryons, which could prevent the emergence of  GRB 
\cite{shemi90}.\\
Here we investigate cases where both components carry different 
spins to see if these baryon free funnels still emerge. We analyze systems 
where one member carries no spin and the 
other star either rotates with the angular frequency of the orbit in the same 
direction as the orbit (spin parallel to the orbital angular momentum),
 opposite to it (antiparallel to the orbit) or with the spin lying in
 the orbital plane (see Fig. \ref{spins}). In each case the spin frequencies
 are set equal to the orbital frequency at the initial separation 
($a_0= 45$ km), corresponding to a spin period of 2.9 ms.

\subsection{Tidal deformation}

When a ns binary system has spiralled down to a distance corresponding
to our initial separation (45 km) the deformation due to mutual tidal forces
will be of the order $\delta R \approx R^4 a_0^{-3} \approx 0.5$ km 
and thus non-negligible. For corotating systems the construction of accurate
numerical initial models is rather straight forward since in a frame corotating
with the binary the fluid velocities vanish. Thus, following Rasio and Shapiro
\cite*{rasio92}, we force the system to relax into a corotating 
equilibrium state by 
applying an artificial friction force proportional to the particle 
velocities. Details of our approach are described in paper I.\\
For the non-corotating binaries we started our calculations with spherical 
stars. These are only very approximate initial conditions and thus
starting with spherical stars leads to oscillations during the inspiral 
phase with a frequency given by 
$\omega_{\rm osc} \approx 2 \pi (G \bar{\rho})^{1/2}$.
In paper I the effects resulting from the oscillations have been investigated 
carefully. They were found to increase the temperatures and internal 
energies of the final configuration, but had a negligible influence on the 
mass distribution. Thus we regard spherical stars at the beginning of our 
calculations as a good approximation for the purposes of this work.\\

\section{Results}

Our calculations start with initial separations $a_{0}= 45$ km and 
follow the last stages of the inspiral and the final coalescence over 12.9 ms.
The characteristics of the investigated cases are summarized 
in Table \ref{runs}. The total energy (after switching off the gravitational 
wave backreaction force) is typically conserved to a few times $10^{-4}$ (see
column 8 in Table 1). 

\subsection{Morphology}

Our runs can be separated into two groups: for the first group (run A - run C)
 the asymmetry of the system is introduced  by individual spins of the stars,
 for the second group
 (run D - run F) it is given by different masses of the components.\\
In run A (where one ns had no spin and the other was spinning parallel to the
 orbital angular momentum) two spiral arms of different size are
formed  after the coalescence. These get wrapped around the central object,
hit its surface supersonically and result in a thick, toroidal, shock-heated 
disk. The final configuration consists of a massive central object, and a 
thick disk surrounded by asymmetrically-distributed rapidly-expanding low 
density material.\\
In case B, where the star with spin rotates against the orbital motion, the 
emerging spiral arms are less well-defined and quickly transform into a 
thick torus of neutron rich material around the central object. In the final
configuration the spiral structure has completely dissolved.\\
In Run C, where the spin of the ns is lying in the orbital plane,
 spiral arms show up for a very short time ($\sim 4$ ms) and 
then transform into a torus engulfing the central object. At the end
of the simulation the central object is surrounded by a thick, dense neutron
torus ($\rho \sim 10^{12}$ - $ \sim 3 \cdot 10^9$ g cm$^{-3}$) that extends
to $\sim$ 100 km which itself is embedded into a extended cloud of low-density
 neutron star debris. 
In this case the material distribution is tilted by a small angle $\varphi$ 
away from the original orbital plane (see Fig. \ref{fun4}).
 This can be explained easily in terms of angular momentum.
If we approximate the rotating star by a rigid, homogeneous  sphere rotating 
with the orbital frequency $\omega$, then we find for our initial conditions
 $L_{\rm spin}$ / $L_{\rm orbit} = 4/5 (R/a)^{2} \approx  4/5 (15 {\rm km}/ 
45 {\rm km})^{2}=4/45$. This translates into an angle of $\varphi= 
\arctan( \frac{L_{\rm spin}} {L_{\rm orbit}}) \approx 5^{\circ}$, which is 
consistent with the numerical result.\\
In the second group (where the masses of the ns are $1.3$ and 1.4 
${\rm M}_{\odot}$) both stars carry 
the same spin: in the direction of the orbital motion 
(since we always use $\omega_{\rm spin} = \omega_{\rm orbit}(a_{0})$
this corresponds to corotation; Run D), both stars without spin (irrotational
configuration; Run E) or both are spinning against the orbital motion (Run F).
In the case of the corotating and  irrotational configurations only one 
spiral arm forms from the material of the less-massive component. In the
 last case (spins against the orbit) no such spiral structure ever emerges 
during the whole evolution. Immediately, a very massive central object, 
engulfed by low density neutron star debris is formed.

\subsection{Mass distribution}

The baryonic masses of the central objects, the tori and the low density 
regions can be found in Table \ref{masses}. 
We use the relation between binding energy and gravitational mass of 
\cite{lattimer89} for an estimate of a lower limit on the 
gravitational mass $\tilde{\rm M}_{\rm co}$ of our central objects:
\begin{equation}
\frac{\tilde{\rm M}_{\rm co}}{{\rm M}_{\odot}}= 5.420 \cdot \left( -1 + \sqrt{1+0.369 \; \frac{\rm M_{b}}{{\rm M}_{\odot}}}\right)\label{gravmass},
\end{equation}
where $\rm M_{b}$ is the baryonic mass. The corresponding values for the 
central objects are given in the third column of Table \ref{masses}.\\
Despite the fact that the observed neutron star masses are all consistent 
with a narrow Gaussian distribution with $m_{ns}= 1.35 \pm 0.04 
{\rm M}_{\odot}$ 
\cite{thorsett99}, present state of the art nuclear EOS
allow for much higher neutron star masses, which could possibly
 indicate that the masses in radio pulsar binary systems might be given 
rather by evolutionary constraints than by nuclear physics.
For example, Weber and Glendenning \cite*{weber92} found maximum masses for rotating 
neutron stars of $\sim 2.5$ ${\rm M}_{\odot}$, most recent  investigations 
\cite{shen98a,akmal98} find upper limits for non-rotating 
stars of $\sim 2.2$ ${\rm M}_{\odot}$ which correspond to the results of Weber 
and Glendenning if a $20 \%$ rotational effect (see e.g. Shapiro and Teukolsky
 \cite*{shapiro83} is added.
The gravitational masses $\tilde{{\rm M}}_{\rm co}$ (see Table \ref{masses})
 for the central objects  are in the critical range between 2.0 and 2.2 
${\rm M}_{\odot}$ (note that the baryonic masses of the initial stars 
are lower (1.3 and 1.4 ${\rm M}_{\odot}$) than in paper I (1.4 and 1.6 
${\rm M}_{\odot}$)). The relativistic stability parameter $a= (J c)/ (G M^2)$
(\cite{stark85}; see column 4 in Table \ref{masses}) is around 0.6 and 
thus substantially lower than the critical value $a_{crit}\approx 1$,
i.e. the increase of the maximum mass resulting from rotational support is
neligible. 
Thus several scenarios are possible. Depending on the maximum neutron star 
mass the central object could collapse immediately after the merger to a black
hole. If the thermal pressure of the hot central object prevents an 
instantanuous collapse to a black hole it could be stabilized on a neutrino
cooling time scale (a few seconds) and then collapse when the thermal 
pressure support is reduced sufficiently. A third possibility is that 
the central object is stable even after the neutrinos have diffused out. The 
collapse could then be triggered by material accreted from the disk on a 
time scale determined by the largely unknown disk viscosity.
It has to be stressed, however, that a secure upper limit on the neutron
star mass from causality arguments \cite{kalogera96} is as high 
as 2.9 ${\rm M}_{\odot}$. 
Since our total baryonic mass is $\le 2.8 {\rm M}_{\odot}$ there is still the 
possibilty that a ns remains stable having accreted the thick disk 
(up to a few tenth of a solar mass) and at least 
those parts of the surrounding low-density material that are not ejected
(see below). This scenario would result in 
supermassive (around twice the ``canonical'' value of 1.4 ${\rm M}_{\odot}$), 
fast rotating, hot neutron stars as an outcome of the coalescence.

\subsection{Kick velocities}

As described in paper I, symmetric systems retain their symmetry with 
respect to the origin during the whole merger evolution. The position of the
 final central object thus coincides with the center 
of mass of the whole mass distribution. Here, however, due to asymmetries
introduced either by different spins or masses of both components, the final 
mass configuration is not symmetric with respect to the origin anymore 
and the central object receives a kick through the ejection of high 
velocity, low density material.\\
 We find large values of $v_{kick}$ for the corotating system with 
different masses (run D; $\sim 600$ km s$^{-1}$), the case where one ns
is spinning in the orbital plane (run C; $\sim 200$ km s$^{-1}$) and  
the system where only one component is fast rotating with the orbital motion
(run A; $\sim 200$ km s$^{-1}$). The other initial configurations 
lead to substantially smaller kicks ($\sim 100$ km s$^{-1}$). 
Thus the merger of asymmetric neutron star binaries may 
result in black holes or supermassive neutron stars with kick 
velocities of several hundred kilometers per second.

\subsection{Baryonic pollution/GRBs}

It was suggested that the funnels that form above the poles of 
the central objects in symmetric systems would be a natural site for a GRB 
fireball to form, since these regions were found to be practically free of
baryons. It has been thought for a long time that the emergence of a GRB
from a region ``contaminated'' with baryons is impossible. If spherical 
symmetry is assumed, an amount as small as $10^{-5}{\rm M}_{\odot}$ 
of baryonic material 
injected into the fireball is enough to prevent a GRB from forming. It was 
only recently that detailed 3D-calculations in the context of the 
collapsar-model \cite{macfadyen99,aloy99} have shown that 
under certain conditions still large relativistic gamma factors can be 
achieved despite a considerable baryonic loading.\\
At present it is still an open question whether the mechanism found for 
the collapsar-model also works in the case neutron star coalescence.
 It is therefore an 
interesting question to ask whether the region above the poles of the 
coalesced object remains free of baryons if the initial neutron star spins
are not alligned with the orbital angular momentum.
To our knowledge only calculations exist where the stellar spins are aligned, 
i.e. parallel or antiparallel, with the orbital angular momentum. It is one 
of the motivations of this investigation to see whether this changes in the 
case of asymmetric systems. Clearly, the stellar spins only contribute 
moderately to the total angular momentum of the system ($\sim 10 \%$, see 
above) and so changing the stellar spins cannot be expected to overturn the 
overall picture of the merger, but in case the collapsar mechanism should not
work even a small amount of baryonic material might be decisive for the fate
of the fireball.\\
In Fig. \ref{fun4} we show density contours  (from $\log(\rho_{[{\rm g cm}
^{-3}]})= 
14.5$ down to 9) in the x-z-plane (i.e. orthogonal to the original orbital 
plane). We find that changing the stellar spins does not endanger the 
formation of the baryon free funnel above the polar caps of the central object.
However, there are indications that in the case of corotation with different
masses  more material is found close to the rotation axis. In the calculation 
of the density contours the smoothing lengths enter and these are in 
SPH generally evolved in a way such that the number of neighbours of 
each particle is kept approximately constant, thus the low density contours
are somewhat biased by the resolution.\\
For the two lower panels of Fig. \ref{fun4} only the particle masses and 
their corresponding positions were used. They are intended to 
give an idea of the particle masses that are contained in a cylinder 
positioned above the pole starting with height $z_{c}$ and radius $r_{c}$. 
Thus a point at ($r_{c}, z_{c}$) on a contour line  of value $c$ indicates 
that a cylinder of radius  $r_{c}$  along the z-axis from $z_{c}$ to infinity 
contains {\em less} than $c$ of baryonic material. These plots reflect that 
more  mass is found close to the rotation axis in the case of a corotating 
system with $q \ne 1$. However, on the one hand these indications are biased
by the current numerical resolution and on the other hand the corotating
case is unlikely to be encountered in reality. Since in the most realistic
case (where the ns have masses 1.3 and 1.4 ${\rm M}_{\odot}$ and no spins;
 run E) the polar regions 
are still free of baryons this place still has to be regarded as an 
excellent site for the emergence of a fireball. 

\subsection{Temperatures}

Low temperatures in the degenerate regime are numerically difficult to 
determine since even the slightest noise in the internal energy density, which
 is our independent variable, can lead to appreciable staggering in the 
temperature. However, this does not influence the dynamical evolution in any
 way. \\
Similar to the symmetric cases (see Ruffert et al. \cite*{ruffert96}, 
paper I) vortex rolls form along the contact interface as soon as the stars 
merge (for a more detailed discussion see paper I). The vortex sheet emerging
at the contact interface is Kelvin-Helmholtz unstable on all wavelengths and
 therefore difficult to model in a 3D calculation (see e.g. Rasio and Shapiro
\cite*{rasio99}). The maximum temperatures of the systems are found in 
these vortices. 
These are highest for the cases with the most shear interaction (spin against
 orbital motion, run F and B) where peak temperatures above $\sim 25$ MeV 
are found (note that due to the lower total mass the mergers are less violent
than the ones described in paper I). The lowest temperatures are found in the
corotating case (run D). One reason is that -due to the initial relaxation-
the system practically does not oscillate during the inspiral. The other 
reason is that this case exhibits the minimum shear motion (zero velocity
in the corotating frame) of all cases.  
The evolution of the peak temperatures of the different cases are shown 
in Fig. \ref{Tmax}.\\
The thick disks are created when the spiral arms are wrapped 
up around the central
object. During this process different parts of the spiral arms collide 
supersonically and their inner parts hit the surface of the central object.
The shock heated disks have in all cases mean temperatures between 3 and 4 MeV.
Due to the shear motion within the disk artificially high temperatures as 
artifacts of the artificial viscosity cannot be fully excluded, not even
with our new viscosity scheme (see appendix) which exhibits a strongly
improved behaviour in shear flows. Thus, these temperatures 
should be regarded as an upper limit to the true disk temperatures.
Our results for the temperatures are close to those reported by Ruffert et al.
 \cite*{ruffert96} for their PPM-calculations.

\subsection{Ejecta}

\begin{table*}[ht]
\caption{Amount of ejected material
(masses are given in solar units).}
\begin{flushleft}
\begin{tabular}{ccccc} \hline \noalign{\smallskip}
run & remark & m$_{\rm ej}$ $(\# \, {\rm part.}), e_i>1$ &  m$_{\rm ej}$ $(\# \, {\rm part.}), E_{\rm i,tot}>0$\\ \hline \\
A   &1.4 \& 1.4 ${\rm M}_{\odot}$, spin: 1& $ 1.49 \cdot10^{-2} \; (134)$ & $1.42 \cdot10^{-2} \; (128)$\\ 
B   &1.4 \& 1.4 ${\rm M}_{\odot}$, spin: 2& $ 2.66 \cdot10^{-3} \; (24) $ & $ 2.66 \cdot10^{-3} \; (24)$\\
C   &1.4 \& 1.4 ${\rm M}_{\odot}$, spin: 3  & $ 1.70 \cdot10^{-2} \; (164)$ & $ 1.74 \cdot10^{-2} \; (168)$\\
D   &1.3 \& 1.4 ${\rm M}_{\odot}$, spin: 4  & $ 3.64 \cdot10^{-2} \; (390)$ & $ 3.62 \cdot10^{-2} \; (387)$\\
E   &1.3 \& 1.4 ${\rm M}_{\odot}$, spin: 5  & $ 1.11 \cdot10^{-2} \; (99) $ & $ 1.14 \cdot10^{-2} \; (101)$\\
F   &1.3 \& 1.4 ${\rm M}_{\odot}$, spin: 6  & $ 2.0  \cdot10^{-4} \; (2)  $ & $ 3.1  \cdot10^{-4} \; (3)  $ \\
G   &1.6 \& 1.6 ${\rm M}_{\odot}$, spin: 4  & $ 1.93 \cdot10^{-2} \; (160)$ & $ 2.07 \cdot10^{-2} \; (172)$  \\
H   &1.6 \& 1.6 ${\rm M}_{\odot}$, spin: 4  & $ 0 \; (0)$ & $ 1.21 \cdot10^{-4}\; (2)$ \\
I   & like run E, std. AV  & $ 1.36 \cdot 10^{-2} \; (134)$ & $ 1.35 \cdot10^{-2}\; (132)$ \\
J   & like run E, no AV    & $ 1.13 \cdot 10^{-2} (98)$ & $ 1.15 \cdot10^{-2}\; (100)$ \\
\end{tabular}
\end{flushleft}
\label{ejectatable}
\end{table*} 

In part the enormous interest in the coalescence scenario of two neutron 
stars arises from its possible importance for nucleosynthesis. Despite
intense research it has not been possible to pin down the astrophysical
 production site of the r-process nuclei.
For their production these nuclei basically need an environment which is 
characterized by the presence of seed nuclei and very high neutron densities.
These conditions are suggestive of explosive events in neutron-rich 
surroundings. The most popular scenario is a type II supernova. 
However, recent calculations 
reveal two severe problems connected with this production site: (i) there is 
no way to produce the observed r-process abundance pattern for nucleon numbers
 below 120 \cite{freiburghaus97,freiburghaus99a} and (ii) the nuclei above 
this value 
can only be reproduced if entropies are applied that exceed the entropy values
 of ``state-of-the-art'' supernova calculations by a factor of 3 to 5
\cite{takahashi94,meyer97c,hoffman97,meyer98,freiburghaus99a}. The only possible way out from
this conclusion would be the introduction of sterile neutrinos (see e.g. 
McLaughlin \cite*{mclaughlin99}).\\
The neutron star merger scenario is a promising r-process
 site  since it  provides in a natural way the neutron 
rich environment needed for the capture reactions.\\
To clarify the importance of the merger scenario for the 
r-process nucleosynthesis 
the following questions have to be answered: (i) how often do such mergers 
occur? (ii) how much mass is (depending of the parameters of the binary system)
ejected per event? and (iii) how much of this material is r-process matter,  
respectively, do we find a solar r-process pattern?
Here we focus on the second point. The third one has been discussed in a 
recent paper \cite{freiburghaus99b}.\\
 Since at the end of our calculations the outermost parts of the neutron star
 debris are basically ballistic, it is a reasonable approximation to treat the
 SPH-particles as point masses in the gravitational potential of the central
 mass distribution. Replacing the central mass by a point mass $M$, we can
 calculate the orbital eccentricities of the point masses $m_{i}$:
\begin{equation}
e_{i}= \sqrt{1+ \frac{2 E_{i} J_{i}^{2}}{G^{2} m_{i}^{3} M^{2}}}, \label{e}
\end{equation}
where $E_{i}$ and $J_{i}$ are energy and angular momentum of particle i.
A particle $i$ is regarded as being unbound if $e_i > 1$. 
Since the assumption of ballistic motion may possibly not be justified
 everywhere
we test the results by calculating the sum of each particle's energies 
(see paper I). The amounts of ejected material according to both criteria are
in good agreement and are given in Table \ref{ejectatable}.\\
 We find that the ejecta for asymmetric coalescences are comparable to those 
from symmetric systems (see paper I). 
The only larger deviation is found for the case where both neutron stars
spin against the orbital motion. Here we can hardly resolve any mass loss 
($\sim 2\cdot 10^{-4}$ M$_{\odot}$) while the corresponding spin 
configuration in paper I ejected $\sim 5\cdot 10^{-3}$ M$_{\odot}$. This
larger value may be a result of spurious entropy generation by the former
artificial viscosity that is now suppressed by our new scheme (see appendix).
However, this configuration is very unlikely to be encountered in nature
and only meant to give a lower limit for the spin dependence of the ejecta. 
To be cautious we have explored the dependence of these results on the AV
with two further test runs: we start from the initial conditions of the 
most probable initial configuration, run E, but use the standard AV scheme
in one case and in the other we use no AV. With the standard AV we find that
$\sim$ 20 \% more material is ejected; without AV the result is practically 
the same as was obtained with our hybrid scheme. Thus, the mass loss
measured in this paper is {\em definitly not an artifact of the AV scheme}.    
The ejecta for the other configurations range from a few times 
$10^{-3}{\rm M}_{\odot}$ to a few times $10^{-2}{\rm M}_{\odot}$
exhibiting a strong sensitivity to the stellar spins.\\
It has to be stressed again that the amount of ejected material is crucially
 dependent on the poorly known nuclear equation of state. 
In test runs using a soft polytrope ($\Gamma= 2.0$) we were not able to resolve
 any mass loss at all (paper I).
To test the sensitivity of the amount of ejecta on the behaviour of the EOS
in different density regimes we  performed two additional test runs. In 
both of these runs we used a polytropic EOS whose adiabatic exponent 
$\Gamma$ varies with density $\rho$:

\begin{equation}
P= (\Gamma(\rho)-1) \rho u,
\end{equation}
where $P$ denotes the pressure and $u$ the specific internal energy. The 
adiabatic exponent was prescribed according to 
\begin{equation}
\Gamma= \Gamma_{1} - (\Gamma_{1}-\Gamma_{2}) \left( \frac{\rho}
{\rho+\rho_{0}} \right).\label{adindex}
\end{equation}
We chose $\rho_{0}= 10^{12}$ g cm$^{-3}$ and took the values from paper I
(2.0 and 2.6) for $\Gamma_{1}$ and $\Gamma_{2}$
(see Fig. \ref{gamma}).
In the first case (run G), the  high density part of matter followed a 
$\Gamma=2.6$-polytrope whereas the low density material was governed 
by a $\Gamma=2.0$-polytrope. In the second case (run H) the polytropic indices
were switched. This allowed us to
start from the initial configuration of runs C and K of paper I 
and thus assured that changes in the amount of ejecta are exclusively due to 
variations in $\Gamma$. Both cases indicate that the results (see Table 
\ref{ejectatable}) are governed by the central part of the merged 
configurations,
i.e. {\em the behaviour of the nuclear equation of state in the high density 
regime basically determines the amount of ejected neutron star debris}. This 
is unfortunately the poorest known regime of the EOS and reveals again that
 tighter limits on the stiffness of the EOS would be highly desirable.
In addition, this sensitivity to the strong field gravitational potential 
in the center of the mass distribution indicates that we are at the limit of
applicability of Newtonian gravity and thus further calculations including
general relativistic gravity are necessary.\\

\section{Summary and discussion}

We have presented Newtonian, 3D calculations of the hydrodynamic 
evolution of  neutron star binary coalescences where we have used the smoothed
 particle hydrodynamics method  coupled to the realistic nuclear equation 
of state of Lattimer and Swesty. Our focus in this investigation was 
on slightly asymmetric binary 
systems, where the asymmetry stemmed either from different masses 
(1.3 and 1.4 ${\rm M}_{\odot}$) or spins of both components (both stars 
of mass 1.4 ${\rm M}_{\odot}$).\\
In all cases a fast rotating central object  with a baryonic mass above 
$2.3 {\rm M}_{\odot}$ formed. Since the exact maximum neutron star mass is 
unknown and our calculations are Newtonian we cannot decide on the fate of 
the central object. It might collapse to a black hole directly
 after the merger, but
it might instead remain stable on a neutrino diffusion or an accretion time 
scale (determined by the largely unknown disk viscosity). Even the final
 creation of supermassive neutron stars with twice the canonical mass 
value of 1.4 ${\rm M}_{\odot}$ cannot firmly be excluded.\\
The central object is surrounded by a thick disk containing 
 between 0.05 and 0.23 ${\rm M}_{\odot}$. Typically a few percent of a solar
mass is in rapidly expanding low density regions. In the cases where this low 
density material is expelled in an asymmetric way, the central object receives
 a kick velocity of several hundred kilometers per second (we found the 
highest kick velocities of $\sim 600$ km s$^{-1}$ for a corotating system 
containing a 1.3 and a 1.4 ${\rm M}_{\odot}$ star), which is comparable 
to the kick velocities of neutron stars that result from asymmetric 
supernova explosions (e.g. Fryer and Kalogera \cite*{fryer97}), 
Fryer et al. \cite*{fryer98}).\\
One motivation to study systems with neutron star spins that are {\em not} 
orthogonal to the orbital plane was to see whether the baryon free funnels
above the poles of the central objects that have been found in previous 
calculations
are ``polluted'' by the injection of baryonic material. However, we did not
observe such an effect.
Funnels similar to the symmetric cases arose for all spin combinations.
 For the cases with the neutron star spins lying in the orbital plane 
the final disk was tilted by a small angle with respect to the original 
orbital plane.\\
As found in earlier investigations (Rasio and Shapiro 1994) even small 
deviations from the mass ratio $q=1$ result in dramatic consequences
 for the overall hydrodynamic evolution of the systems. Here we have 
investigated systems of 1.3  and 1.4 ${\rm M}_{\odot}$ ($q\approx 0.93$). 
The heavier star was
 rather unaffected by the coalescence whereas the lighter one was
 disrupted, forming a layer of debris around the heavier and providing the 
material for the formation of the low density regions (spiral arms, disks). 
For initial corotation parts of this debris are driven towards the rotation 
axis. However, since corotating ns binaries are very unlikely and in 
the more probable spin configurations baryon free funnels appear (to within
the numerical resolution) we still regard the poles above the central object
to be a very promising site for the emergence of a GRB fireball.\\
The amounts of material that become unbound during the coalescence of a
slightly asymmetric neutron star system are comparable to those found in 
previous calculations for symmetric systems ($\sim 10^{-2} {\rm M}_{\odot}$ 
for our most realistic configuration with 1.3 and 1.4 ${\rm M}_{\odot}$,
no spins and the LS-EOS). However, several uncertainties concerning the amount
of ejected material per event enter.
\begin{enumerate}
\item {\em Artificial viscosity} (AV): The standard form of the 
artificial viscosity tensor (e.g. Monaghan 1992) is known to create 
spurious entropy in shear flows. This is of no concern for strictly 
corotating systems (zero fluid velocity in a corotating frame) but for 
other spin configurations a shear layer forms at contact. 
Thus one might suspect that some particles are artificially ejected through
spurious forces.
 However, the AV scheme used in the presented 
calculations  largely suppresses spurious shear forces (in a 
differentially rotating star the viscous time scales are increased by 
two orders of magnitude, see appendix) while still keeping the ability 
to resolve shocks properly. In test calculations without AV for the most probable configuration (non-spinning ns of masses 1.3 and 1.4 ${\rm M}_{\odot}$)
it turned out that the amount of ejected material ($10^{-2} {\rm M}_{\odot}$)
is definitely {\em no artifact of the AV scheme}.

\item {\em Gravity}: The perhaps major shortcoming of the current investigation
is the use of Newtonian gravity. It is expected that the deeper general
relativistic potentials render the escape of low density material more
difficult. Recent Post-Newtonian SPH-calculations (Ayal et al. 1999, 
Faber and Rasio 1999) support this view. These efforts towards strong-field 
gravity are undoubtedly steps in the right direction, but it is not 
immediately obvious how
much closer to reality these approaches are since the PN-expansion parameters
are $\sim M R^{-1} \approx 0.2$ at the neutron star surface and thus 
questionable (Ayal et al. used stars of $\sim 0.5 {\rm M}_{\odot}$, Faber and 
Rasio introduced ad hoc terms to increase the stability of their schemes).\\
A further approximation is the use of a point mass backreaction accounting
for forces arising from the emission of gravitational waves. It has to be 
switched off shortly before the merger. The emission of gravitational waves,
however, will still be significant for a short time after the stars first touch
(both Ayal et al. and Faber and Rasio find secondary peaks in the gravity wave
luminosities) and thus slightly more angular momentum may be radiated away.
This might influence the amount of ejecta as well. 

\item {\em EOS}:  
The amount of material ejected is extremely 
sensitive to the nuclear equation of state of the neutron stars. In Rosswog 
et al. (1999) we found a strong dependence on the stiffness  of the EOS (e.g.
$1.5 \cdot 10^{-2} {\rm M}_{\odot}$ for $\Gamma=2.6$ compared to no resolvable
 mass loss for $\Gamma=2.0$). Thus we performed test runs with a 
pseudo-polytropic EOS whose stiffness varied with density. They 
revealed that the amount of ejecta is basically determined by the behaviour of
 the nuclear EOS in the high-density regime. This is problematic in two 
respects. First, the considerable uncertainties in the behaviour of
nuclear matter at supranuclear densities constrain our knowledge concerning 
the ejected material and second, it indicates that the strong-field region in 
the centre influences the ejecta which are found far away. This might lead
one to question the applicability of Newtonian gravity here.
\end{enumerate}
While the first two points (AV, gravity) argue for lower values of 
the ejected material the last one could decrease (for a softer EOS) as 
well as increase (stiffer EOS) the amount of ejecta.\\
Arzoumanian \cite*{arzoumanian99} argue that pulsar ages derived from 
spin down 
time scales are generally overestimates and thus birth and merger rates 
of ns-ns systems are underestimated. They place a firm upper limit of 
$10^{-4} {\rm yr}^{-1} {\rm galaxy}^{-1}$ on the ns-ns merger rate, which
 agrees with the value estimated by Tutukov and Yungelson \cite*{tutukov93} 
(for a careful discussion of merger rates and their uncertainties see
 \cite{fryer99}). With this upper limit even an amount as small as 
$10^{-4}{\rm M}_{\odot}$ per event might contribute substantially to the 
enrichment of the universe with heavy element material (see Fig. 26 in
Rosswog et al. 1999).\\
Concerning the abundance distributions in the ejecta it has to be stressed 
that fully dynamical r-process calculations using a realistic EOS, a careful 
account of weak interactions (including neutrino transport) and a consistent
 coupling  with the hydrodynamic evolution are still lacking. We have 
performed dynamical  r-process network calculations, including heating 
processes due to the decay of unstable heavy nuclei, that follow the expansion
rates of our hydrodynamic calculations (Freiburghaus et al. 1999b; preliminary
results are given in Rosswog et al. \cite*{rosswog98c}). All of the 
ejected material is found to undergo the r-process. 
If the initial $Y_{e}$ is too low ($\sim 0.05$)
the resulting abundance pattern looks s-process like, for $Y_{e}= 0.08 - 0.15$
the solar r-process pattern above the A=130 peak is excellently reproduced,
while below this peak non-solar (under-abundant) patterns are found.
Thus if the initial $Y_e$ of the ejecta is in the right range we predict
 underabundant r-process patterns for nuclei with
A $<$ 130 in very old, metal-poor stars, while the nuclei above $A=130$ 
should be found with abundances  close to the solar pattern.
Actually, there seems to be support for this view coming from observation
\cite{wasserburg96,cowan99}. Since the neutron star merger rate is
 significantly lower than the core collapse
supernova rate, the merger would have to eject more material per event to
explain the observations. This would lead to some kind of ``clumpyness''
 in the early distribution of r-process material. Such a clumpyness is
actually observed \cite{sneden96}: the relative abundances in very old
stellar populations match the solar pattern very well, but their absolute
values show large variations in different locations.\\
The main problem for type II supernova ejecta with its high  $Y_{e}$ values to 
reproduce the solar r-process pattern stems from the too high entropies that 
are needed. However, this obstacle might be overcome in the collapsar model. 
Perhaps here the required entropies could be attained and, depending on the
 rate and the ejecta per event, an interesting amount of r-process material 
could be synthesized.
However, this r-process scenario might run into problems with the observations
 of old, metal poor stars due to the short life times of its progenitors 
and/or the large observed r/Fe ratios reaching values of three times solar.
These observations seem to indicate that the emergence of r-process material 
is delayed with respect to iron \cite{mcwilliam95,mcwilliam97}.
This delay would disfavour the collapsar model since it is only consistent 
with low mass SN-progenitors \cite{mathews92}, but could find 
a natural explanation due to the delay caused by the inspiral of a ns-ns 
binary.

  \acknowledgements
  This work was  supported by the Swiss National 
			  Science Foundation under grants No. 20-47252.96 and 
			  2000-53798.98 and in part by the National Science 
			  Foundation under 
			  Grant No. PHY 94-07194 (S.R., F.-K.T.), M.B.D. 
			  acknowledges the support of the 
			  Royal Society through a University
			  Research Fellowship,
			  T.P. was supported by the US-Israel BSF grant 95-328.

  \begin{appendix}

\section{Artificial viscosity}

Like other schemes in numerical hydrodynamics SPH uses an artificial
viscosity (AV) to resolve shocks properly.
The basic task of AV is to keep particles from interpenetrating each other
and to ensure the correct solution of the energy equation across the 
shock front. If the kinetic energy of matter passing through the shock
is not correctly transformed into heat, unphysical oscillations are encountered
in the post-shock region.
The ``standard'' SPH viscosity \cite{monaghan92} allows for an accurate 
resolution of shock fronts (smoothed over $\sim 3$ smoothing lengths) and 
damps out post-shock oscillations.
However, since the corresponding AV-tensor (see below) 
contains terms $\vec{r}_{ij} \cdot \vec{v}_{ij}$, i.e. the inner 
product of the position and velocity difference vectors of particles 
$i$ and $j$, spurious forces may be introduced in the case of pure shear flows.
SPH has often been criticized for unphysical effects introduced in this way.\\
Our goal here is to find an AV prescription that is able to resolve shocks
properly where necessary, but reduces viscous forces as far as possible in 
shear flow situations. For a better control of artificial viscosity we 
suggest here a hybrid scheme that benefits from two recent improvements 
of the AV treatment in SPH: the introduction of time dependent viscosity 
parameters $\alpha$ and $\beta$ \cite{morris97} and the so-called 
Balsara-switch \cite{balsara95}
 to suppress viscous forces in pure shear flows.
The  modified artificial viscosity tensor reads:

 \[ \Pi_{ij} = \left\{ \begin{array}{lr}
\frac{- \alpha_{ij} (t) \; c_{ij} \; \mu_{ij} \;  + \;\beta_{ij} (t)\; 
\mu_{ij}^{2}}{\rho_{ij}} \quad \quad & \vec{r}_{ij} \cdot \vec{v}_{ij} \le 0 \\
\quad \quad \quad \quad 0  & \; \vec{r}_{ij} \cdot \vec{v}_{ij} > 0  ,
\end{array}    \right. \]
where $\beta_{ij}= 2 \cdot \alpha_{ij}$, $\alpha_{ij}= \frac{\alpha_{i}
+\alpha_{j}}{2}$, $c_{ij}= \frac{c_{i} + c_{j}}{2}$, $h_{ij}= \frac{h_{i} 
+ h_{j}}{2}$.
 The $c_{k}$ denote the particle sound velocities, $h_{k}$ the smoothing 
lengths, and  $\vec{r}_{k}$ and $\vec{v}_{k}$ positions and velocities,
 $\vec{r}_{ij}=\vec{r}_{i}-\vec{r}_{j}$, $\vec{v}_{ij}=
\vec{v}_{i}-\vec{v}_{j}$.
The term that is responsible for possibly spurious effects 
is modified by the Balsara-switch:
\begin{equation}
\mu_{ij} = \frac{h_{ij}\vec{r}_{ij} \cdot \vec{v}_{ij} }{r_{ij}^{2}+\eta 
h_{ij}^{2}} \cdot \frac{f_i+f_j}{2},
\end{equation}
where $\eta= 0.01$ and 
\begin{equation}
f_i= \frac{|\nabla \cdot \vec v|_i}{|\nabla \cdot \vec v|_i + 
|\nabla \times \vec v|_i + \eta' c_i/h_i}.
\end{equation}
The SPH-prescription for the velocity curl is:
\begin{equation}
(\nabla \times \vec v)_i = \frac{1}{\rho_i} \sum_j m_j \vec{v}_{ij}
\times \nabla_i W_{ij}
\end{equation}
and the velocity divergence is given by
\begin{equation}
(\nabla \cdot \vec v)_i = - \frac{1}{\rho_i} \sum_j m_j \vec{v}_{ij}
\cdot \nabla_i W_{ij}
\end{equation}
In a pure shock the divergence term will dominate over the curl, 
$|\nabla \cdot \vec v|_i >> |\nabla \times \vec v|_i$, and thus 
$f_i \rightarrow 1$, reproducing the standard viscosity term. In pure shear
flows, however, the curl dominates, $|\nabla \times \vec v|_i 
>> |\nabla \cdot \vec v|_i$, and thus strongly suppresses viscous terms since
$f_i \rightarrow 0$. To prevent numerical divergences $\eta ' (=10^{-4})$ is
introduced.\\
 To have enough viscosity where it is needed, i.e. to resolve shocks properly, but to keep it on a 
minimal level otherwise the viscosity parameters $\alpha$ and 
$\beta (=2\alpha)$  are 
allowed to evolve in time. This is realized by determining the viscosity 
coefficient $\alpha_{i}$ from an additional equation that has to be 
integrated together with the other dynamic equations. It contains a 
term that drives the decay towards minimum values $\alpha_{min}$ on a 
time scale $\tau_i$ and a source term $S_i$ responsible for the rise of 
the parameters in the presence of shocks:

\begin{equation}
\frac{d \alpha_{i}}{dt}= - \frac{\alpha_{i}- \alpha_{min}}{\tau_{i}} + S_{i}.
\end{equation}

To keep the viscosity parameters in a well-defined intervall and to allow for
sufficiently {\em fast} rise if a shock is detected, we have slightly 
modified the original source term equation:
\begin{equation}
S_{i}= max(-( \vec{\nabla} \cdot \vec{v})_{i} (\alpha_{max}-\alpha_{i}),0),
\label{viscsource}
\end{equation}
After a number of numerical experiments we have chosen the following 
parameters whose appropriateness will be shown in the subsequent tests:
 $\alpha_{max}= 1.5$, $\alpha_{min}=0.05$
and $\tau_{i}= \frac{h_{i}}{\epsilon \; c_{i}}$ with $ \epsilon  = 0.2$.\\
To validate this form of AV we compare its capabilities to the standard scheme
in the context of three test cases:
(i)   a 1D shock tube to test the ability to resolve shocks, 
(ii)  a stationary, differentially rotating star where viscous time scales
      are calculated to quantify the amount of spurious shear forces and 
(iii) a close, tidally interacting non-equilibrium binary neutron star 
system that is driven to the dynamical instability limit by viscosity.\\

For the shock tube test we start with the initial conditions described 
in \cite{hernquist89} (see also references therein):
\begin{eqnarray}
\rho &=& 1,    \quad \quad \; P = 1, \quad  \quad \quad \; v = 0 \quad 
\mbox{for } x < 0 \nonumber \\
\rho &=& 0.25, \quad P = 0.1795, \quad v = 0 \quad \mbox{for } x \ge 0 
\label{stsu}
\end{eqnarray}
with a ratio of specific heats $\gamma= 1.4$ and 1000 equal mass particles with
$m_i= 0.75/1000$ initially distributed in a way that (\ref{stsu}) is satisfied.
Smoothing lengths are fixed to $h_i=0.0065$. 
In Fig. \ref{st} we show a comparison of the density 
and velocity profiles at $t=0.15$ between the standard AV with $\alpha= 1$
and $\beta=2$ (often suggested as ``standard'' values, 
see e.g. Monaghan 1992) and our hybrid scheme. 
The hybrid scheme shows a slightly sharper resolution of the shock front, 
 apart from that the calculated shock-profiles are almost indistinguishable. 
The time dependent viscosity parameter $\alpha$ is also displayed in the 
left panel. It is sharply peaked at the shock front where it reaches values
of $\approx 0.6$ and is close to the minimum value of 0.05 otherwise. 
In summary, the hybrid scheme exhibits shock resolution capabilities
very similar to the standard scheme, but the average value of the 
viscosity parameter $\alpha$ is largely reduced.\\
In a second, 3D, test we will focus on shear flows in a configuration 
specific to our merger application. 
To this end we construct a stationary, 
differentially rotating star in a way similar to Lombardi et al. 
\cite{lombardi99}: (i) we start from a relaxed spherical star obeying the
LS-EOS, (ii) give all the particles a constant velocity $v_0= 0.1 c$, i.e.
the star rotates differentially with $\omega (r) \sim r_{cyl}^{-1}$, where
$r_{cyl}$ is the distance to the rotation axis, (iii) to reach a rotating 
equilibrium state, unwanted velocities are damped out by applying 
an artificial drag force $\vec{f}_d \sim (v_0 \cdot \hat{e}_{\phi} 
- \vec{v}_{i})$.
A star relaxed in this way exhibits a $\omega$-profile close to 
$\sim r_{cyl}^{-1}$, only near the origin, where the kernels
overlap, the  $\omega$-singularity is smoothed to finite values.
In a totally inviscid star, the viscous times scales 
$\tau_{visc,i}= v_i / \dot{v}_{i,visc} = v_i/(|\sum_j m_j \Pi_{ij}
\nabla_i W_{ij}|) $ should be infinite.
However, by means of the terms in the artificial viscosity tensor that contain
 $\vec{v}_{ij} \cdot \vec{r}_{ij}$
spurious viscous forces are introduced in pure shear flows that lead to
 finite viscous time scales. Thus, the aim of an improved viscosity scheme
has to be the increase of these $\tau_{visc,i}$. 
For a differentially rotating equilibrium configuration with 5000 particles
constructed in the three steps described above
the viscous time scale $\tau_i$ of each particle is shown in Fig. \ref{tvisc}
 as a function of the distance to the rotation axis $r_{cyl}$. Plus signs 
refer to the standard viscosity ($\alpha= 1, \beta=2$), triangles to a time 
dependent scheme ($\alpha_{max}=1.5, \alpha_{min}= 0.05, c= 0.2$), and 
circles to the hybrid scheme (time dependent $\alpha$ and Balsara switch).
The introduction of time dependent viscosity parameters alone increases the
viscous times (for the chosen parameter set) by approximately one order of 
magnitude. The additional Balsara switch further improves the time scales by 
another order of magnitude.\\
In our last, 3D test we start out from a close binary system where both stars
have 1.4  ${\rm M}_{\odot}$, an initial separation $a_0= 45$ km and do not 
posess any spin. Since the initial stars are spherical and the system is close
 enough for tidal interaction the 
binary components start to oscillate and thereby transform orbital energy 
into oscillatory energy. Since we are interested here in (spurious) effects
of viscosity, the initial system is not provided with radial velocities, no
gravitational backreaction force is applied during this calculation. The 
inspiral and final merger is triggered by viscosity alone. In Fig. \ref{rcm}
the evolution of the center of mass distance of the binary is shown (10000
particles). For reasons of comparison the evolution of a system
with initial radial velocities and backreaction force is also displayed. The 
introduction of time dependent AV parameters substantially delays the time
until the system becomes dynamically unstable, the additional Balsara-factor
further retards the coalescence. 

\end{appendix}

\bibliography{literat,aamnem99}
\bibliographystyle{aabib99}

\clearpage


\clearpage


\begin{figure}[h]
\psfig{file=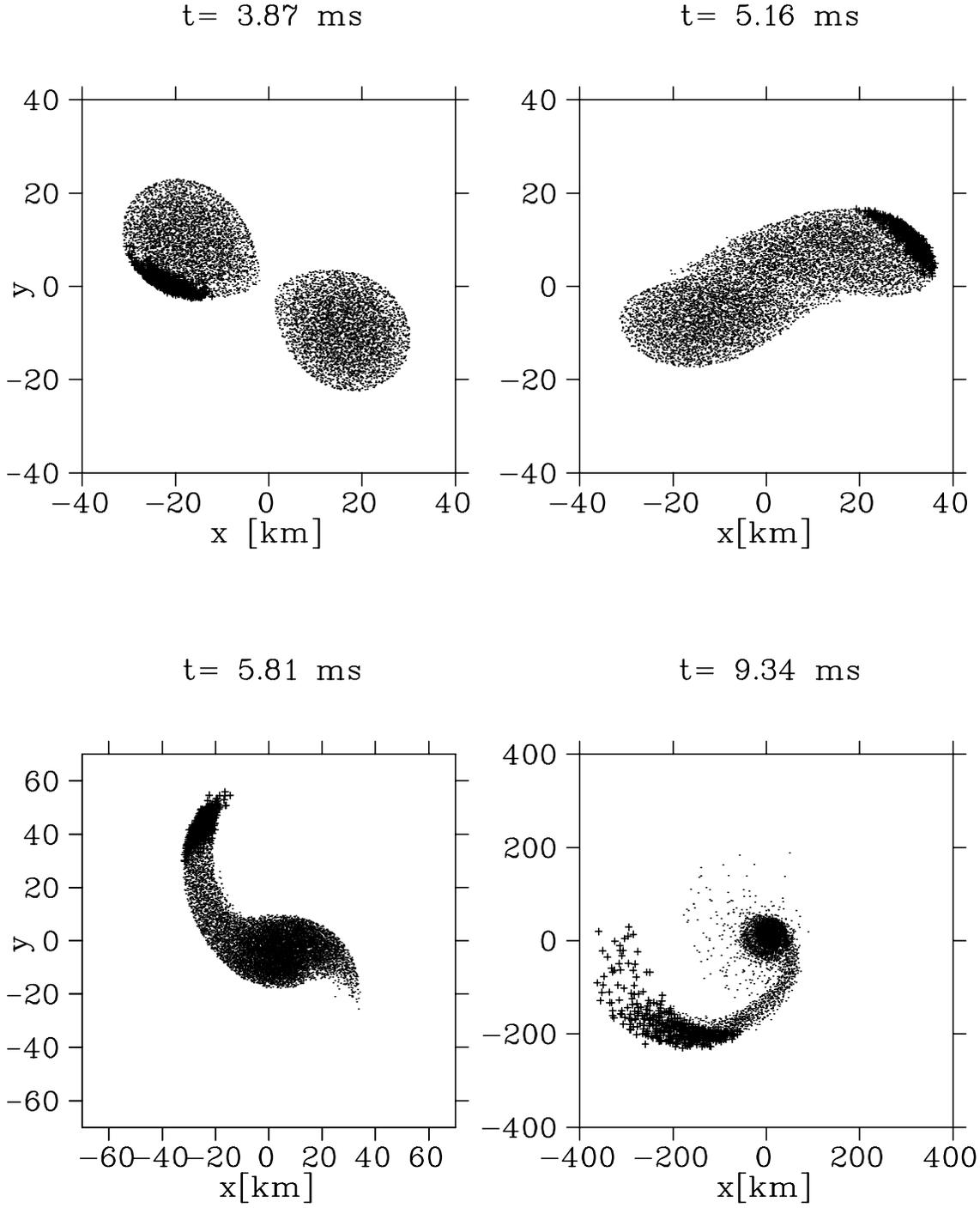,width=15cm}
\caption{\label{cr1314}Morphology of a corotating binary system with
neutron stars of masses 1.3
 and 1.4 ${\rm M}_{\odot}$ (run F). Dots indicate projections of 
particle positions onto the orbital plane, crosses show particles that 
are unbound at the end of the simulation.}
\end{figure}\clearpage


\begin{figure}[h]
\psfig{file=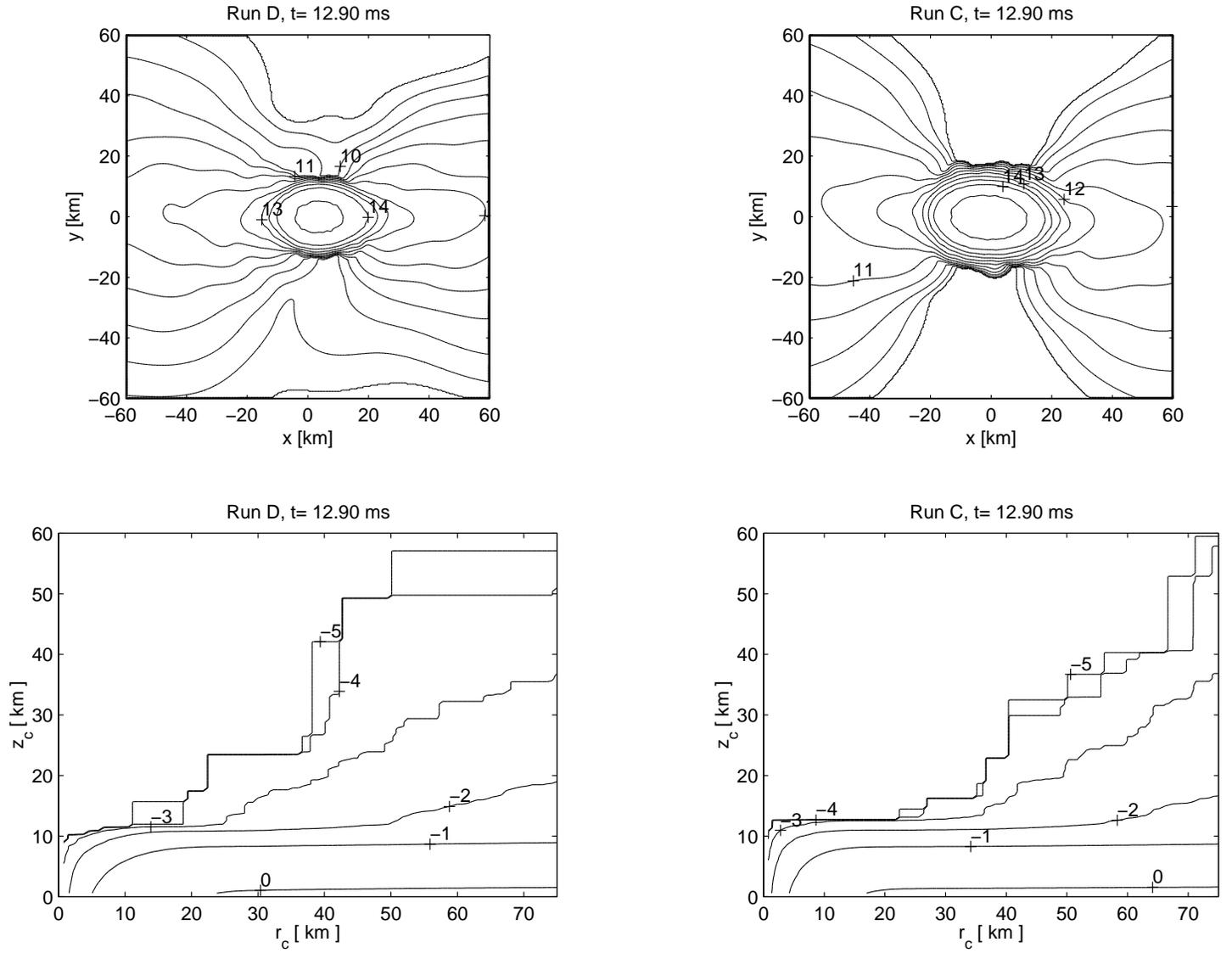,width=15cm,angle=90}
\caption{\label{fun4} Cut through xz-plane (at the end of the calculation)
of the corotating system (run D; left column) and the one where one spins lies
in the orbital plane (run C; right column). The labels in the first line
of plots indicate the logarithm of the density contours (in g cm$^{-3}$).
In the lower plots a point at ($r_{c}, z_{c}$) on a
 contour line  of value $c$ indicates that a cylinder of radius  $r_{c}$ 
 along the z-axis from $z_{c}$ to infinity contains less than $c$ of 
 baryonic material.}
\end{figure}\clearpage

\begin{figure}[h]
\psfig{file=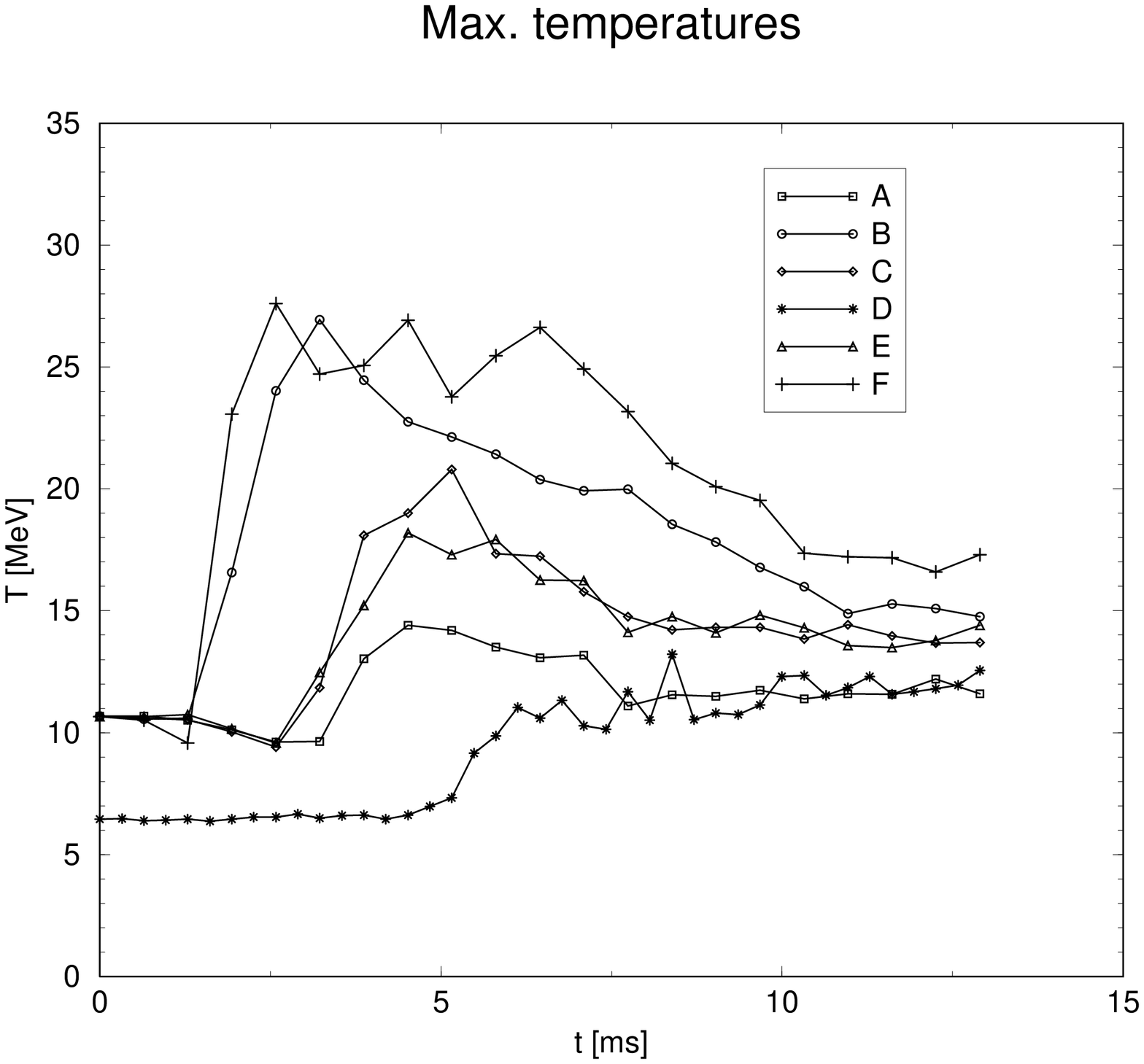,width=14cm,angle=-90}
\caption{\label{Tmax} Maximum temperatures of the different runs.}
\end{figure}\clearpage

\begin{figure}[h]
\psfig{file=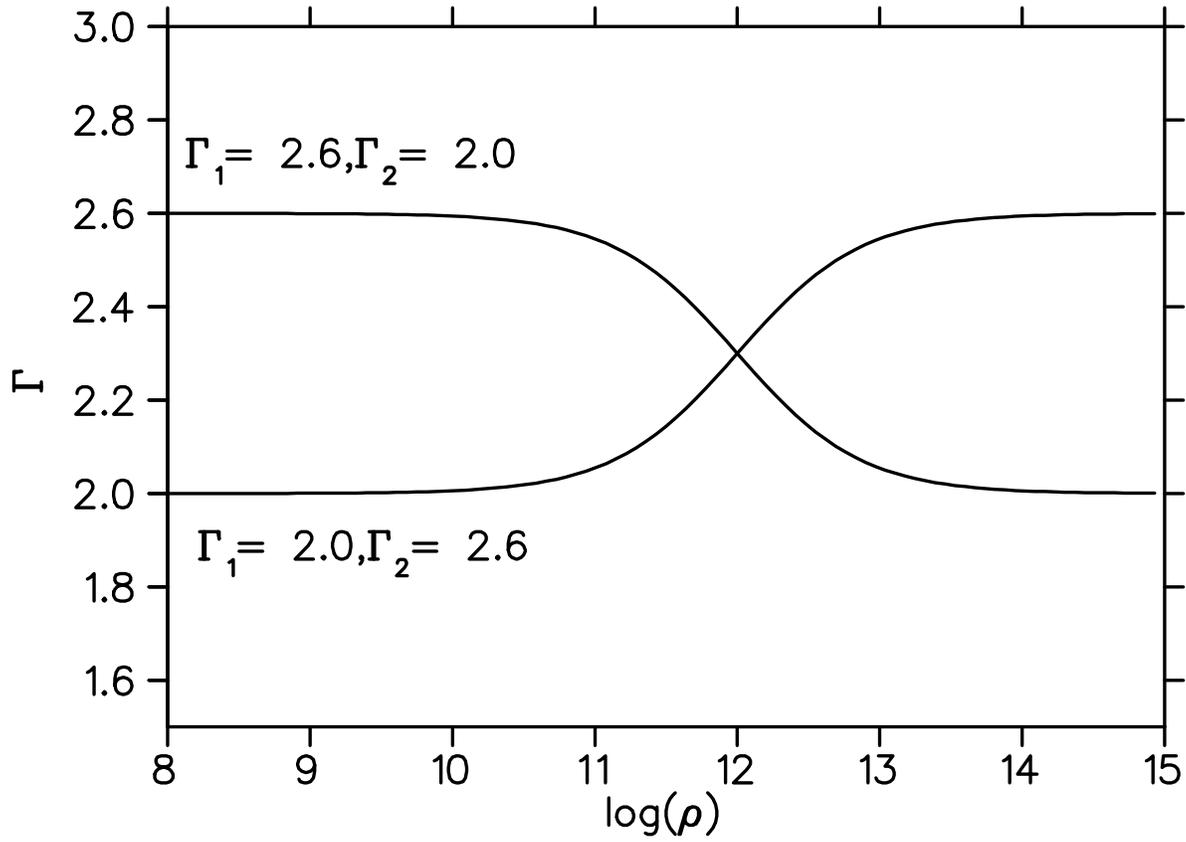,width=11cm,angle=90}
\caption{\label{gamma} Dependence of the adiabatic index on density for 
the pseudo-polytropic EOS used in the test runs H and J.}
\end{figure}\clearpage


\begin{figure}[h]
\epsfig{file=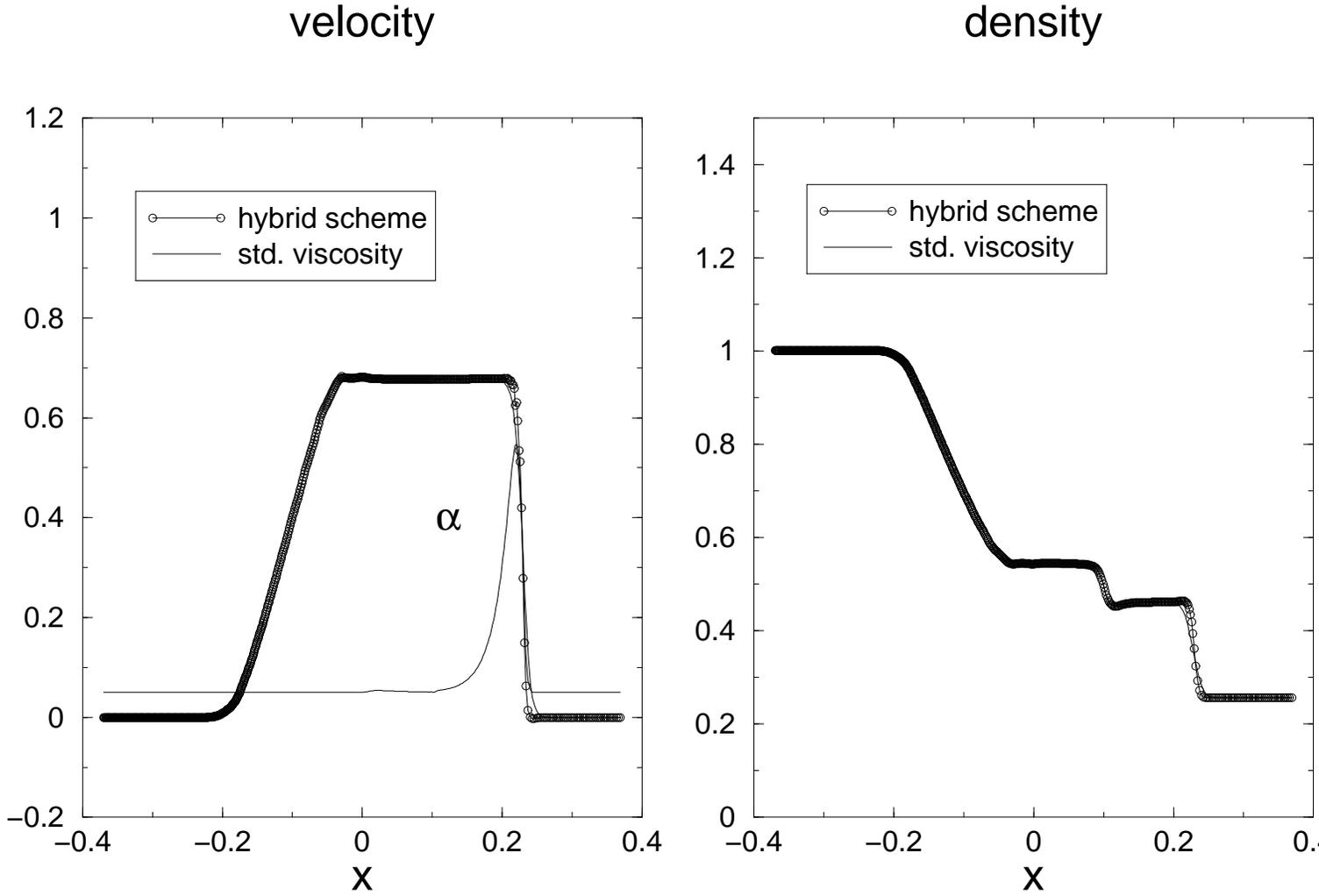,width=14cm,angle=-90}
\caption{\label{st} Comparison of shock tube results at $t=0.15$ between the
standard SPH artificial viscosity ($\alpha=1, \beta=2$) and our hybrid AV 
scheme ($\alpha_{max}=1.5, \alpha_{min}= 0.05, c= 0.2$). The time dependent 
viscosity parameter $\alpha$ is also shown in the left panel.}
\end{figure}\clearpage


\begin{figure}[h]
\epsfig{file=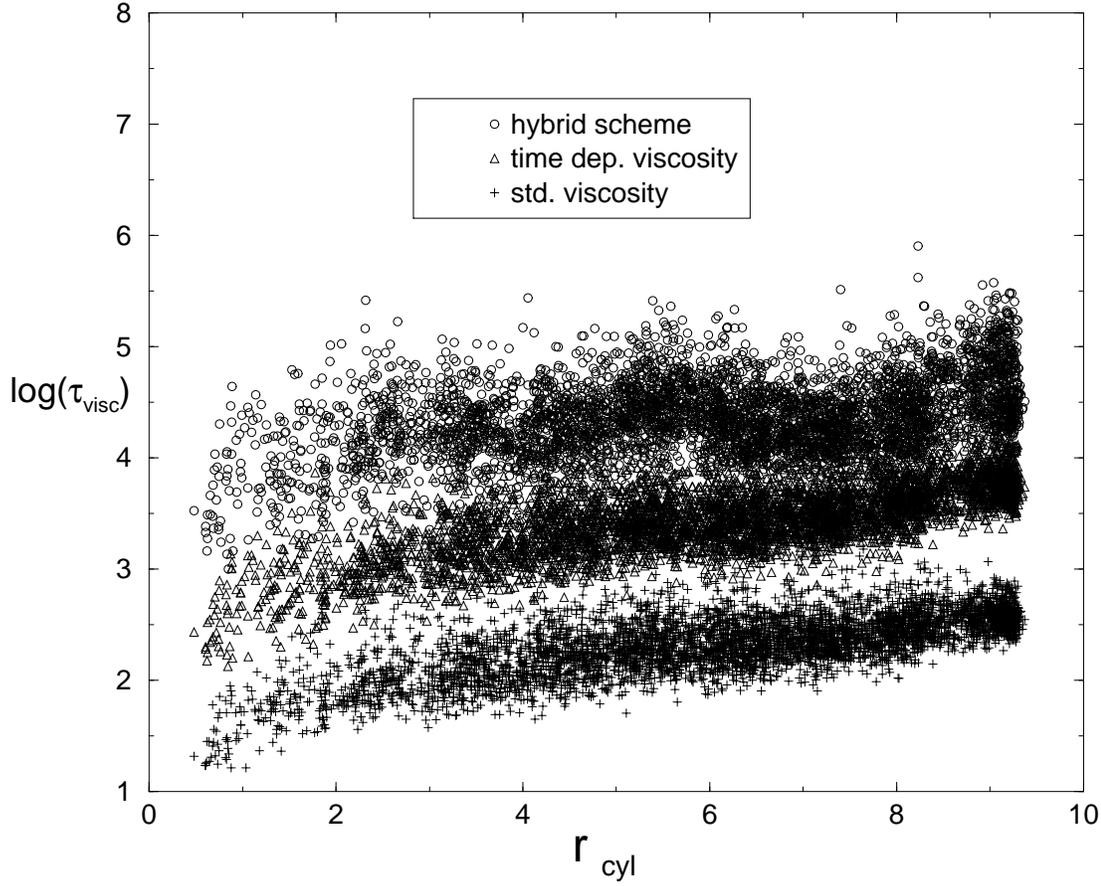,width=12cm,angle=-90}
\caption{\label{tvisc}Viscous time scales $\tau_{visc,i}= v_i / 
\dot{v}_{i,visc} = v_i/(|\sum_j m_j \Pi_{ij} \nabla_i W_{ij}|)$ in a 
differentially rotating equilibrium configuration for the standard AV, the
time dependent scheme of Morris and Monaghan ($\alpha_{max}=1.5, 
\alpha_{min}= 0.05, c= 0.2$) and the hybrid scheme used here. All quantities 
are given in code units ($G=c={\rm M}_{\odot}=1$).}
\end{figure}\clearpage


\begin{figure}[h]
\epsfig{file=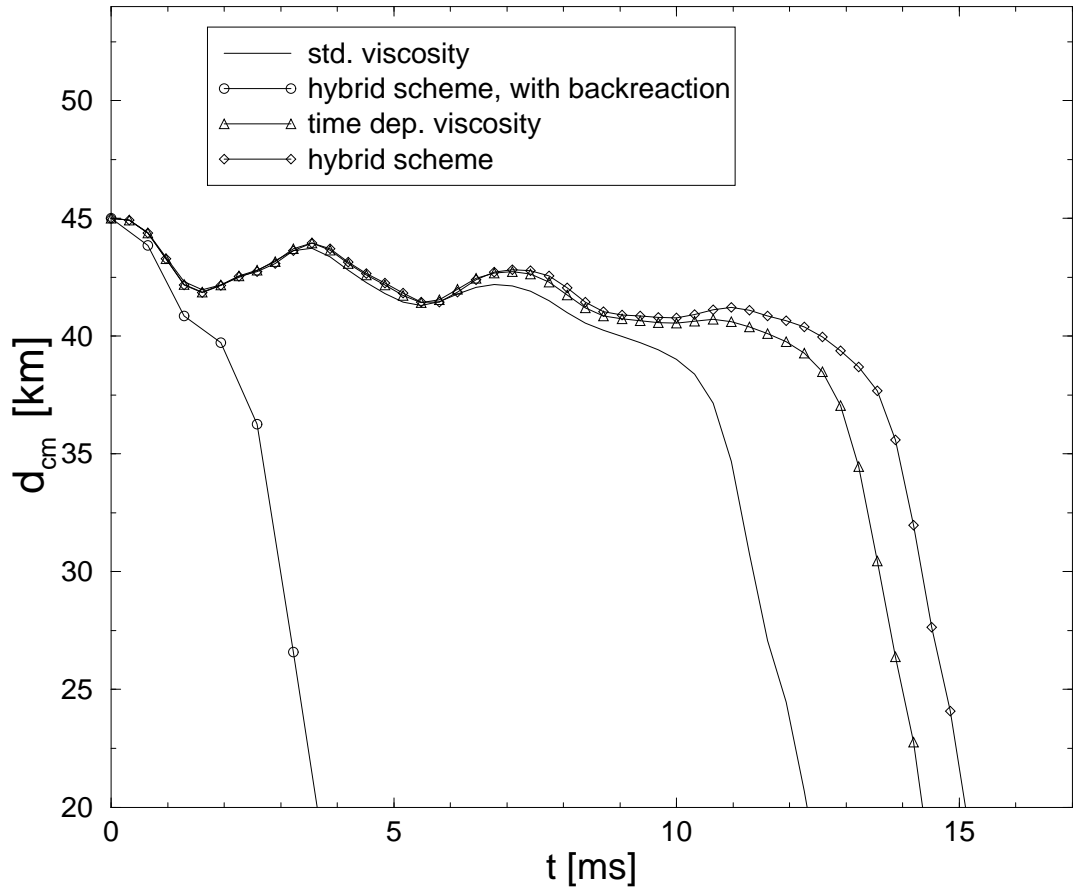,width=12cm,angle=-90}
\caption{\label{rcm} Evolution of the distance between the centers of mass
of an irrotational sytem (both stars with 1.4  ${\rm M}_{\odot}$). Apart
from one test case (circles) the inspiral is entirely driven by viscosity.}
\end{figure}

\end{document}